\documentclass[apj,twocolumn,twocolappendix]{aastex63}

\newcommand{\Ri}{R_{\rm{in}}}
\newcommand{\etatens}{\overline{\eta}}
\newcommand{\alphatens}{ {\overline{\alpha}}_{\rm dyn} }
\newcommand{\Cs}{c_{\rm s}}
\renewcommand{\vec}[1]{\mathbf{#1}}
\newcommand{\tens}[1]{\mathsf{#1}}
\newcommand{\pd}[2]{\frac{\partial #1}{\partial #2}}
\newcommand{\DS}{\displaystyle}

\usepackage{ctable} 
\usepackage{epstopdf}
\usepackage{pbox}
\usepackage{gensymb}
\usepackage{amsmath}
\usepackage{amssymb}

\begin{document}
\title{MHD accretion-ejection: jets launched by a non-isotropic accretion disk dynamo. II.

       A dynamo tensor defined by the disk Coriolis number}

\author[0000-0003-1454-6226]{Giancarlo Mattia}
\altaffiliation{Member of the International Max Planck Research School for Astronomy \& Cosmic Physics at the University of Heidelberg}

\author[0000-0002-3528-7625]{Christian Fendt}

\affiliation{Max Planck Institute for Astronomy, Heidelberg, Germany}

\correspondingauthor{Giancarlo Mattia}
\email{mattia@mpia.de, fendt@mpia.de}


\begin{abstract}
Astrophysical jets are launched from strongly magnetized systems that host an accretion disk surrounding a central object. 
Here we address the question how to generate the accretion disk magnetization and field structure required for jet launching.
We continue our work from Paper I \citep{Mattia2020}
considering a non-scalar accretion disk mean-field $\alpha^2\Omega$-dynamo in the context of large scale disk-jet simulations.
We now investigate a disk dynamo that follows analytical solutions of mean-field dynamo theory, 
essentially based only on a single parameter, the Coriolis number.
We thereby confirm the anisotropy of the dynamo tensor acting in accretion disks, allowing to relate both the
resistivity and mean-field dynamo to the disk turbulence.
Our new model recovers previous simulations applying a purely radial initial field, while allowing for a more 
stable evolution for seed fields with a vertical component.
We also present correlations between the strength of the disk dynamo coefficients and the dynamical 
parameters of the jet that is launched, and discuss their implication for observed jet quantities.
\end{abstract}


\section{Introduction}
Astrophysical jets are launched from a wide range of astrophysical objects such as young stellar objects (YSO), micro-quasars or 
active galactic nuclei (AGNs).
It is commonly accepted that these jets are launched from strongly magnetized systems that host an accretion disk 
surrounding a central object \citep{2014prpl.conf..451F, 2015SSRv..191..441H, 2019FrASS...6...54P}.
Further agreement is on the key role of the large-scale magnetic field for the jet acceleration and collimation.

As we have further detailed in Paper I,
the origin of the jet-launching disk magnetic field is still not completely understood.
A promising model scenario that can provide such a large-scale disk magnetic field is that of an 
accretion disk dynamo process.

Essentially, astrophysical dynamos are thought to be of turbulent, thus small-scale origin.
On the other hand, one is interested in the dynamical effects of the generated
magnetic field on these systems on the large scales, i.e. the whole disk jet system.
The disk turbulence -- providing both a turbulent dynamo effect and also a turbulent magnetic diffusivity -- is 
generally thought to be generated by the magneto-rotational instability, MRI \citep{1991ApJ...376..214B}.

Given by the nature of the problem -- the combination of small-scale effects of turbulence, and the need for a
large-scale, thus global disk magnetic flux --
two paths of modelling the dynamo effect have been pursued.
These are (i) direct simulations, that study the natural amplification of the magnetic field
by the turbulent dynamics of the medium 
(see, e.g., 
\citealt{2010MNRAS.405...41G, 2013ApJ...767...30B}),
and (ii) the so-called mean-field approach 
(see, e.g., 
\citealt{1980opp..bookR....K,1995A&A...298..934R, 2001A&A...370..635B}),
by which an mean electro-magnetic force is derived from averaging the turbulent motions of the medium
that under certain conditions may give rise to a dynamo effect amplifying a weak seed magnetic field
(for further references we refer to our introduction in Paper I).
The mean-field dynamo is usually designated as $\alpha^2\Omega$-dynamo, where the $\alpha$ stands for the field
amplification (poloidal and toroidal field) by the turbulence, while the $\Omega$ stands for the induction of the
(toroidal) magnetic field by differential rotation.

In our work we follow the second approach.
In Paper I we have applied various (ad-hoc) choices for the three components of the dynamo tensor, $\alpha_\phi, \alpha_\theta, \alpha_R$.
We had found that the toroidal magnetic field component is always amplified by the turbulent dynamo 
component $\alpha_\phi$ and by the $\Omega-$effect.
The component $\alpha_\phi$ is strongly correlated to the amplification of the poloidal magnetic 
field, such that a stronger $\alpha_\phi$ results in a more magnetized disk, which then launches a faster, more massive
and more collimated jet.

In contrast, the amplification of the poloidal field depends substantially on the existence 
of dynamo-inefficient zones, which, subsequently, affect the overall jet-disk evolution, thus accretion and ejection.
We found that not only a stronger dynamo component $\alpha_\theta$ but also a non isotropic radial component $\alpha_R$, leads to the formation of dynamo-inefficient zones.
It became clear that the formation of the dynamo-inefficient zones can also be triggered by a vertical component of the 
initial magnetic field, even for a weak dynamo component $\alpha_\theta$.
A strong $\alpha_\theta$ component triggers the formation of the dynamo-inefficient zoned predominantly in the inner
disk region.

Here, in Paper II, we expand on this, investigating an analytical model of turbulent dynamo theory \citep{1995A&A...298..934R} 
that incorporates both the magnetic diffusivity and the turbulent dynamo term, connecting their strength and their 
amount of anisotropy by only one parameter, the Coriolis number $\Omega^*$.

This paper is organized as follows. 
In Section \ref{sec:model} we summarize the main features of our numerical setup, while for an extended presentation
we refer to Paper I.
In Section \ref{sec:Rudiger} we introduce the standard accretion disk dynamo model and we apply it to large scale disk-jet 
simulations.
We summarize our paper in Section \ref{Sec:conclusions}.
In the Appendix we present a resolution study demonstrating the quality of our approach.

%
\section{Model approach}
\label{sec:model}
We solve the time-dependent, resistive MHD equations applying
the PLUTO code \citep{2007ApJS..170..228M} version 4.3, 
on a spherical grid $(R,\theta,\phi)$ assuming axisymmetry.
We refer to $(r,z,\phi)$ as cylindrical coordinates.

We have further detailed our model approach in Paper I.
Here, for convenience, we provide a summary of the most essential points.
The resistive, time-dependent MHD equations, considering a mean-field dynamo are,
\begin{equation}
    \begin{array}{l}
    \DS\pd{\rho}{t}   +  \nabla\cdot(\rho\vec{v}) = 0 \\ \noalign{\medskip}
    \DS\pd{\rho\vec{v}}{t}  +  \nabla\cdot\left[\rho\vec{v}\vec{v}
    + \left(P + \DS\frac{\vec{B}\cdot\vec{B}}{2}\right)\tens{I} 
    - \vec{B}\vec{B}\right] = -
    \rho\nabla\Phi_{\rm g} \\ \noalign{\medskip}
    \DS \pd{e}{t}  +  \nabla\cdot\left[\left(e + P + \DS\frac{\vec{B}\cdot\vec{B}}{2}\right)\vec{v}
              - (\vec{v}\cdot\vec{B}) \vec{B}
              + \etatens\vec{J}\times\vec{B}\right] 
              = \Lambda_{\rm{cool}} \\ \noalign{\medskip}
    \DS\pd{\vec{B}}{t}  + \nabla\times(\vec{B}\times \vec{v} -
        \alphatens \vec{B} + \etatens\vec{J}) = 0,
    \end{array}
\end{equation}
where the primitive variables ($\rho$,$\vec{v}$,$p$,$\vec{B}$) are, respectively, the gas density, velocity and pressure and the magnetic field, while $e$, whose dependence on $\rho$ and $p$ is defined in the equation of state, is the internal energy.
The tensors $\alphatens$ and $\etatens$ describe the $\alpha$-effect of the mean-field dynamo and the magnetic diffusivity.
As in Paper I, for the sake of simplicity, we set the cooling term to be equal to the ohmic heating.

The length and time scales, as the MHD primitive variables, are normalized to their 
value at the inner disk radius $R_{\rm in}$ (e.g. the time unit is given as 
$t_{\rm{in}} = \Ri/v_{\rm{K,in}}$).

The computational domain has a range of $R=[1,100]\Ri$ in the radial direction, where a stretched grid is applied, and a range of $\theta=[10^{-8},\pi/2-10^{-8}]\simeq[0,\pi/2]$ in the angular direction, where a uniform grid is applied.
The numerical resolution is  $[N_R\times N_\theta] = [512\times128]$ grid cells, 
which allows to resolve the initial disk height $H = 0.2 r$ with 16 cells.

For the resolution study (see Appendix \ref{sec::resolution}) we have applied a resolution of 
$[N_R\times N_\theta] = [1024\times256]$ and $[N_R\times N_\theta] = [256\times64]$ grid cells, namely 32
and 8 cells per disk height, respectively.

As the MHD equations are scale-free, our normalized variables can be scaled to a variety of jet sources.
We apply the same scaling as in previous works (see Tab.1 in Paper I).

The numerical algorithms are piecewise parabolic interpolation method (PPM, see \citealt{2014JCoPh.270..784M}) 
for the spatial integration, a third-order Runge-Kutta scheme for the time integration and a 
Harten-Lax-van Leer (HLL) Riemann solver \citep{2009book.123..123}.
We apply the method of Upwind Constrained Transport (UCT, \citealt{2004JCoPh.195...17L}) in order to preserve 
the divergence of the magnetic field.
We choose a Courant-Friedrichs-Lewy time stepping with  $CLF=0.4<1/\sqrt{\rm{N_{dim}}}$.

\subsection{Initial and boundary conditions}
The initial and boundary conditions are identical to those of Paper I.
Here we summarize them for convenience.

The initial state of the disk structure is obtained as a solution of the hydrostatic equilibrium, assuming self-similarity,
and neglecting the weak initial seed magnetic field.
Thus, as given by the initial self-similarity, every (initial) characteristic speed will scale as the Keplerian velocity, 
$\propto R^{-1/2}$.
We assume a polytropic gas, $P\propto\rho^\gamma$.
We set the ratio between the isothermal sound speed and the Keplerian velocity at the disk mid-plane of the inner 
radius to be $\epsilon=\Cs/v_\phi\left|_{\theta=\pi/2}\right. = 0.1$.

Outside the disk we define a hydrostatic corona,
\begin{equation}
    \rho_{\rm c} = 
    \rho_{\rm{c,in}} R^{1/(1-\gamma)},
     \,\,\,\,
     P_{\rm c}=\frac{\gamma-1}{\gamma}\rho_{\rm{c,in}}
    R^{\gamma/(1-\gamma)},
\end{equation}
with $\rho_{\rm{c,in}}=10^{-3}\rho_{\rm in}$.
If not specified otherwise, we set the initial magnetic field as purely radial and exponentially 
decreasing in vertical direction, with a maximum plasma-beta of $10^{-5}$.

The boundary conditions are identical to Paper I and are summarized in Table \ref{tab::boundaries}. 
Along the rotational axis and the equatorial plane the standard symmetry conditions are applied.
The inner radial boundary is divided into two different areas, 
considering (i) the disk accretion ($\theta > \pi/2-2\epsilon$), and 
the (ii) coronal area ($\theta < \pi/2-2\epsilon$).
Across the inner and outer boundaries both the density and the pressure are extrapolated by a power law.

\begin{table*}
\caption{Boundary conditions \label{tab::boundaries} }
 \centering
\begin{tabular}{ccccccccc}
 \hline
  & $\rho$ & $p$ & $v_R$ & $v_\theta$ & $v_\phi$ & $B_R$ & $B_\theta$ & $B_\phi$ \\
 \hline 
 Inner disk   & $\propto R^{-3/2}$ & $\propto R^{-5/2}$ & $\propto R^{-5/2}\leqslant 0$ & 0       & $\propto R^{-1/2}$ & Slope                 & Slope                  & $\propto R^{-1}$ \\
 Inner corona & $\propto R^{-3/2}$ & $\propto R^{-5/2}$ &                     $0.2$ & $0$       & $\propto R^{-1/2}$ & Slope                      & Slope & 0 \\
 Outer disk   & $\propto R^{-3/2}$ & $\propto R^{-5/2}$ & Outflow$\leqslant0$          & Outflow & Outflow            & $\nabla\cdot\vec{B}=0$ & $\propto R^{-1}$       & $\propto R^{-1}$ \\
 Outer corona & $\propto R^{-3/2}$ & $\propto R^{-5/2}$ & Outflow$\geqslant0$          & Outflow & Outflow            & $\nabla\cdot\vec{B}=0$ & $\propto R^{-1}$       & $\propto R^{-1}$ \\
 \hline
\end{tabular}
\end{table*}

Along the inner coronal boundary we prescribe $B_\phi = 0$, while we  
adopt a power-law for the boundary area towards the inner disk and along the outer boundary.
For $v_\phi$ we prescribe a power law across the inner boundary, while the standard PLUTO outflow (zero gradient) 
conditions are applied along the outer boundaries for all three velocity components ($v_R$,$v_\theta$ and $v_\phi$).
We also require the radial velocity to be non-positive at the outer disk boundaries and non-negative at the outer 
coronal boundaries.
The component $v_\theta$ is set to 0 at the inner boundary, while the radial component follows a power law in the 
inner disk boundary
and a weak inflow into the domain of $v_R = 0.2$ along the coronal boundaries. 

Since using a constrained transport, only the component $B_\theta$ needs to be defined,
while $B_R$ is recovered by solenoidal condition. 
Across the outer boundaries $B_\theta$ to follow a power law, while at the inner boundaries we prescribe 
the poloidal magnetic field inclination, choosing an angle
\begin{equation}
    \varphi = 70\degree\left[1+\exp\left(-\frac{\theta-45\degree}{15\degree}\right)\right]^{-1},
\end{equation}
where $\varphi$ is the angle between the magnetic field and the initial disk surface.
The radial component of the magnetic field is then computed by the code through the divergence-free condition 
of the magnetic field.

\subsection{The model for diffusivity and dynamo}
For a thin disk, the non-diagonal components of the mean-field dynamo and the magnetic diffusivity tensors are negligible.
We apply a dynamo tensor derived in the theoritecal analysis by  
\citet{1995A&A...298..934R,2000A&A...353..813R,2001A&A...370..635B},
\begin{equation}
\label{eq::dynamo}
    \alphatens=(\alpha_R,\alpha_\theta,\alpha_\phi) = -\overline{\alpha}_0 \Cs F_\alpha(z),
\end{equation}
where $\Cs$ is the adiabatic sound speed at the disk mid-plane and $F_\alpha(z)$ is a profile function,
\begin{equation}
    F_\alpha(z) = \left\{\begin{array}{ll}
    \sin\left(\pi\DS\frac{z}{H}\right) & z \leq H \\ \noalign{\medskip}
    0 & z > H
    \end{array}\right.
\end{equation}
with $H$ being the initial disk pressure scale height, that confines the dynamo action within the accretion disk.
For a thin disk, the non-diagonal components of the mean-field dynamo tensor are negligible.

For the magnetic diffusivity tensor we adopt an $\alpha$-prescription,
\begin{equation}
\label{eq::diffusivity}
    \etatens = (\eta_R,\eta_\theta,\eta_\phi)=\overline{\eta}_0\alpha_{\rm{ss}}\Cs H F_\eta(z),
\end{equation}
where $\alpha_{\rm{ss}}$ is the dimensionless parameter of turbulence \citep{1973A&A....24..337S}.
The profile function that confines the diffusivity within the disk region is
\begin{equation}
    F_\eta(z) = 
    \left\{\begin{array}{ll}
      1 & z\leq H \\ \noalign{\medskip}
      \exp\left[-2\left(\DS\frac{z-H}{H}\right)^2\right] & z > H
    \end{array}\right.
\end{equation}
Here, as in Paper I, we 
apply the so-called {\em strong diffusivity} model that we have previously invented 
\citep{2014ApJ...793...31S,2014ApJ...796...29S},
\begin{equation}
\label{eq::ssm}
    \alpha_{\rm{ss}} = \sqrt{\frac{2}{\gamma}}\left(\frac{\mu_D}{\mu_0}\right)^2,
\end{equation}
with $\mu_0 = 0.01$ and
$\mu_{\rm D}$ being defined as the ratio between the {\em average} total magnetic field (vertically averaged at a certain radius) 
in the disk and the gas pressure at the disk mid-plane \citep{2014ApJ...796...29S}.
As demonstrated in \citep{2014ApJ...796...29S,2018ApJ...855..130F}, this approach allows to perform a stable evolution of the disk-jet structure over very long simulation times (up to 500.000 inner disk rotations).

\subsection{Dynamo number and dynamo quenching}
We define a dynamo number as in Paper I,
\begin{equation}
\label{eq::dynnum}
    {\cal D} =\DS\frac{\alpha_\phi\Omega H^3}{\eta_{\rm{disk}}^2}
\end{equation}
\citep{1995A&A...298..934R,2000A&A...353..813R}.
where the quantity $\alpha_\phi$ is computed at $z = H/2$.

This definition of the dynamo number is the product of the azimuthal magnetic Reynolds number,
${\cal R}_\Omega=|\Delta\Omega|H^2/\eta_{\rm{disk}}$, based on the shear of the flow $\Delta\Omega$,
and the magnetic Reynolds number ${\cal R}_\alpha = \alpha_\phi H / \eta_{\rm{disk}}$, based on the $\alpha$-effect 
(considering $\alpha_\phi$ as the strongest dynamo contribution in disks).

For a diffusivity profile almost constant with radius, the dynamo number $\cal D$ would scale almost 
linearly with the radius.
However, this is just a first estimate, since the disk diffusivity does not follow a constant profile.
Moreover, the disk orbital velocity and the mid-plane sound speed do undergo small changes through the temporal evolution.
The dynamo number also strongly depends on the turbulent viscosity $\alpha_{\rm ss}$.
We thus expect, as the magnetic diffusivity grows because of the strong diffusivity model, the dynamo number to decrease to a sub-critical value at which the amplification of the magnetic field fades.

For the quenching of the dynamo effect we apply the model of \citep{2014ApJ...796...29S}.
This basically involves quenching by diffusivity, through the strong feedback of the disk magnetization on the 
magnetic diffusivity.
The study of physically more self-consistent feedback models for dynamo quenching will be subject of our future work.

In Paper I we have elaborated that the dynamo number is not always a useful parameter characterizing 
the mode of amplification of the magnetic field.
In particular, we have found that the initial critical dynamo number depends on several factors e.g. the 
number  of grid cells or the magnetic field configuration.
For this reason an initial critical dynamo number is very hard to find and it may not follow an unambiguous prescription
(see also \citealt{1988ApJ...331..416S,1990ApJ...362..318S,1994A&A...283..677T}).

This holds in particular in the presence of a dynamo inefficient zone.
The dynamo number has a strong dependence on the disk diffusivity, while the dynamo inefficient zones are characterized by a low diffusivity (compared to the rest of the accretion disk).
Therefore, the dynamo number is an useful characteristics to spot dynamo inefficient zones within the accretion disk, 
but here it does not imply any amplification of the magnetic field within such zones.

Nevertheless, the dynamo number still remains a key parameter in order to understand the evolution and saturation of the dynamo action.

\section{Simulations of an accretion disk dynamo}
\label{sec:Rudiger}
In Paper I we have considered a an-isotropic mean-field dynamo tensor as a toy model for a realistic 
accretion disk dynamo.
In this section we put this on more physical grounds, considering a dynamo tensor that follows from analytical
dynamo theory.
In particular, we now model the magnetic diffusivity $\etatens$ and the mean-field dynamo $\alphatens$ by applying 
the mean-field theory of \cite{1995A&A...298..934R,2001A&A...370..635B}.
Here, the strength and distribution of the tensor components of both diffusivity and dynamo are constrained by the 
mean-field theory of turbulence.

The basic assumptions made are that the accretion disk is sufficiently ionized and that the effects of rotation on
turbulence can be described by the Coriolis number
\begin{equation}
  \Omega^* = 2\Omega\tau_c
\end{equation}  
where $\Omega$ is the basic rotation frequency and $\tau_c$ is the turbulence correlation time.

The latter variable cannot be recovered from large scale simulations, and it is a key parameter in order to connect 
the disk scale and the turbulent time and length scales.
Direct simulations (see e.g. \citealt{2010MNRAS.405...41G}) have recovered a typical magnitude of $\Omega^*\simeq0.4$, 
but in order to explain the amplitude of the dynamo this value might be larger by an order of magnitude.
For this reason we will present a parameter study of $\Omega^*$ in Section \ref{sec::param_mag}.

\begin{figure*}
\centering
\includegraphics[width=0.48\textwidth]{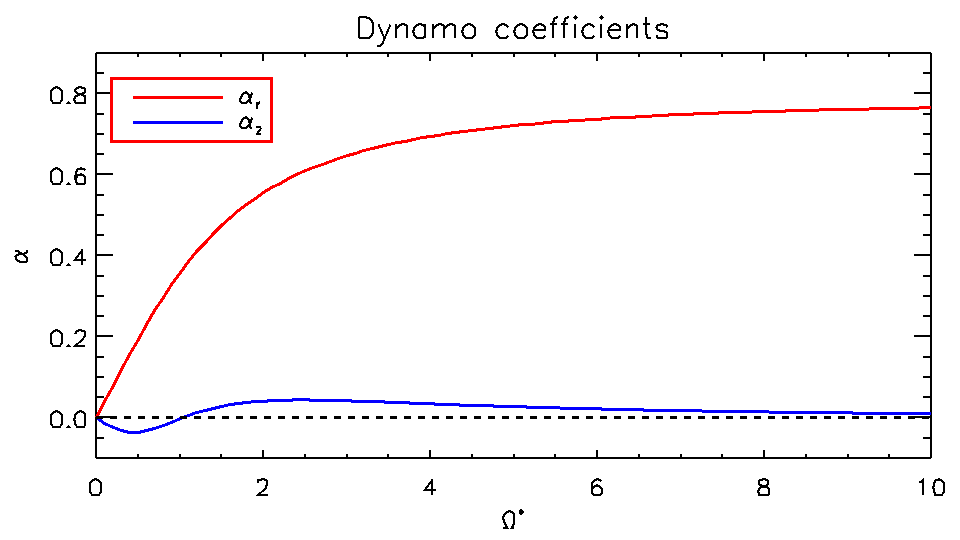}%
\includegraphics[width=0.48\textwidth]{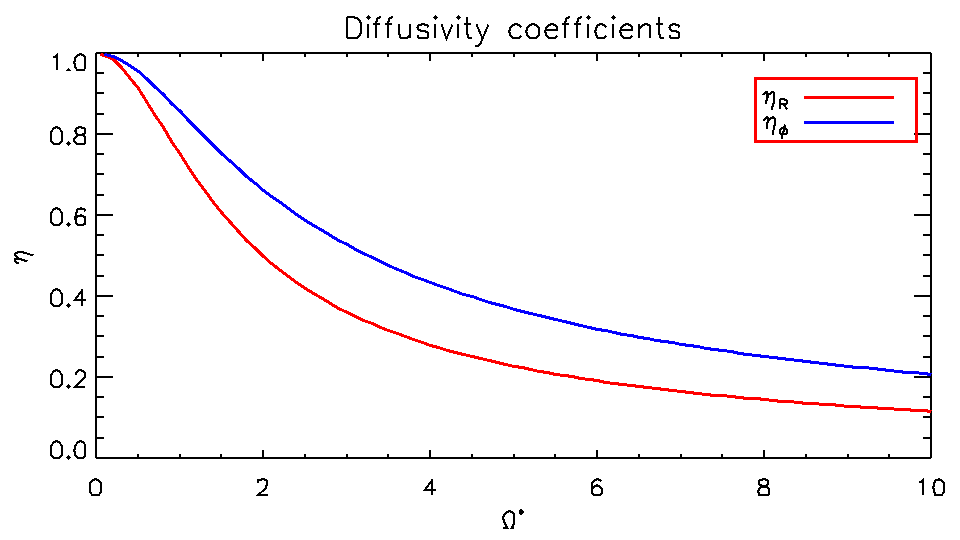}
\caption{Diagonal components of the dynamo tensor (left), $\alpha_r$ and $\alpha_z$, 
              and the magnetic diffusivity tensor (right), $\eta_R$ and $\eta_\phi$,
for different Coriolis numbers $\Omega^*$.
}
\label{fig::alpha_coeff}
\end{figure*}

%
%
%
\subsection{The $\alpha$-tensor}
An essential assumption for the  $\alpha$-tensor is that we are considering a {\em thin disk}.
In this case, the non-diagonal components of the dynamo tensor are negligible \citep{2001A&A...370..635B}.
The explicit form of the dynamo term we have described by Eq.~\ref{eq::dynamo}.
The strength of the respective components of $\alpha$ tensor in cylindrical coordinates is 
\begin{equation}
\label{eq::alphacoeff}
    \left\{\begin{array}{lcl}
    \alpha_{0,r} & =  & \DS\frac{1}{2\Omega^{*3}}\left(\Omega^2 + 6
    -\DS\frac{6+3\Omega^{*2} - \Omega^{*4}}{\Omega^*}\arctan\Omega^*\right) \\ 
      \noalign{\medskip}
      \alpha_{0,z} & = & \DS\frac{1}{2\Omega^{*3}}\left(-\DS\frac{10\Omega^{*2}+12}{1+\Omega^{*2}}
      + \DS\frac{2\Omega^{*2}+12}{\Omega^*}\arctan\Omega^*\right)\\ \noalign{\medskip}
      \alpha_{0,\phi} & = & \alpha_r,
    \end{array}\right..
\end{equation}
\citep{1995A&A...298..934R}.
These component are plotted in the left panel of Fig.\ref{fig::alpha_coeff}.
We notice that for larger $\Omega^*$ the horizontal component $\alpha_r$ overcomes the vertical component $\alpha_z$.
Moreover, the vertical component changes sign around $\Omega^*\simeq1.0$.

While the tensors for the alpha dynamo and the magnetic diffusivity are given in various forms
(compare Equations~\ref{eq::alphacoeff} and \ref{eq:eta-tens}), 
we have transformed all tensor components to the spherical coordinate system we apply for all the simulations discussed 
here (since the dynamo equations of \citet{1995A&A...298..934R} is given in cylindrical coordinates).
So, once the cylindrical components of the dynamo vector are computed, they are rotated in order to recover 
the components also in the spherical coordinates.

%
%
%
\subsection{The diffusivity model}
The magnetic diffusivity tensor follows the same general structure as the dynamo tensor (diagonal, and therefore 
treated as a vector).
For the time evolution of the diffusivity, we again adopt the model described in Eqs.~\ref{eq::diffusivity} and
\ref{eq::ssm}.
However, the quantity $\overline{\eta}_0$ which determines the strength and the anisotropy of the diffusivity tensor, 
is computed following \citet{1995A&A...298..934R},
\begin{equation}
    \left\{\begin{array}{lcl}
      \eta_{0,R} & = & \DS\frac{3}{4\Omega^{*2}}\left[1+\left(\DS\frac{\Omega^{*2}-1}{\Omega^*}\right)\arctan\Omega^*\right] \\ \noalign{\medskip}
      \eta_{0,\theta} & = & \eta_{0,R} \\ \noalign{\medskip}
      \eta_{0,\phi} & = & \DS\frac{3}{2\Omega^{*2}}\left[-1+\left(\DS\frac{\Omega^{*2}+1}{\Omega^*}\right)\arctan\Omega^*\right]
    \end{array}\right.
    \label{eq:eta-tens}
\end{equation}

We note that, contrary to the dynamo prescription, the magnetic diffusivity is computed directly in spherical coordinates.
The reason is the way the $\eta_\parallel$ and $\eta_\perp$ are computed in \citet{1995A&A...298..934R}.
The latter can be directly transformed in spherical coordinates, while the dynamo is computed in cylindrical coordinates.
However, in the thin disk approximation (which is the case of this paper), the spherical and cylindrical components are
only little different. 

The right panel of Fig.~\ref{fig::alpha_coeff} shows the different components of the magnetic diffusivity as a function of 
the Coriolis number $\Omega^*$.
If the turbulence is weak, $\Omega^*<1$, the magnetic diffusivity is basically isotropic \citep{1995A&A...298..934R}.
For strong turbulence, the diffusivity becomes highly anisotropic.
Overall, the turbulence has a major impact on both the dynamo action and the diffusivity.
We point out that the ratio between $\eta_\phi$ and $\eta_R$ in the limit of fast rotation and high turbulence 
($\Omega^*\simeq10$) is comparable with the one used previously \citep{2014ApJ...793...31S,2014ApJ...796...29S}. 

%
%
%
\subsection{A reference simulation}
\label{sec::ref}
The main aim of this paper is to investigate jet launching by a mean-field dynamo 
based on a {\em physical} model of dynamo theory \citep{1995A&A...298..934R}.
In our new approach, the parameter which governs both the mean-field dynamo and the magnetic diffusivity is the 
Coriolis parameter $\Omega^*$. 
We will discuss below simulations applying different Coriolis numbers in the range $\Omega^*\in[0,10]$, 
therefore changing the strength of the dynamo and the diffusivity.

For a reference simulation we have chosen a Coriolis number of $\Omega^* = 10$, while the other parameters (see above)
are taken from \citet{2018ApJ...855..130F}.
Our reference simulation is mainly used to provide a link to the toy models discussed above and that prescribe 
certain combinations of the dynamo tensor.
With the present section we therefore also link the toy model to the physical theory of \citet{1995A&A...298..934R}

A Coriolis number $\Omega^* = 10$
may be considered as high \citep{2010MNRAS.405...41G,2015ApJ...810...59G}, this magnitude has commonly been
used for example of studies of a direct dynamo \citep{1995A&A...298..934R,2000A&A...353..813R} in order to describe
rotating disks for which turbulence has a major effect on the mean-field dynamo.

The run time of our reference simulation (denoted as {\em OM10} from now on) is $t_{\rm F} = 10000$,
corresponding to $\simeq 1500$ inner disk rotations. 
This time is needed to reach a quasi-steady state across the majority of the domain.
As for \citet{2014ApJ...796...29S}, this time is not dictated by numerical issues, 
but chosen in order to save CPU time, as the configuration of the accretion-ejection system does not really change afterwards.


\begin{figure*}
\centering
\includegraphics[width=0.24\textwidth]{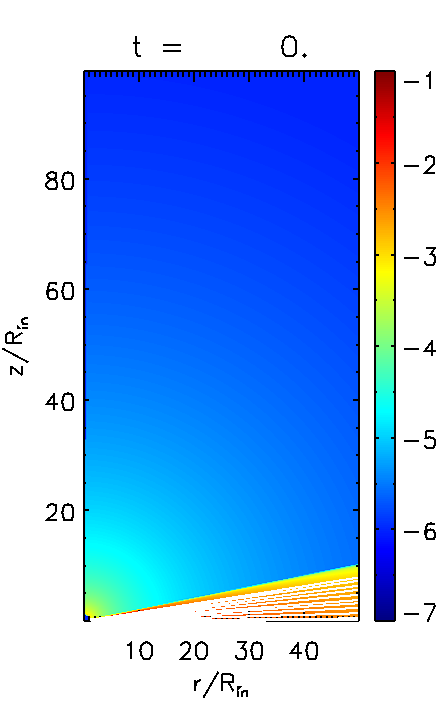}%
\includegraphics[width=0.24\textwidth]{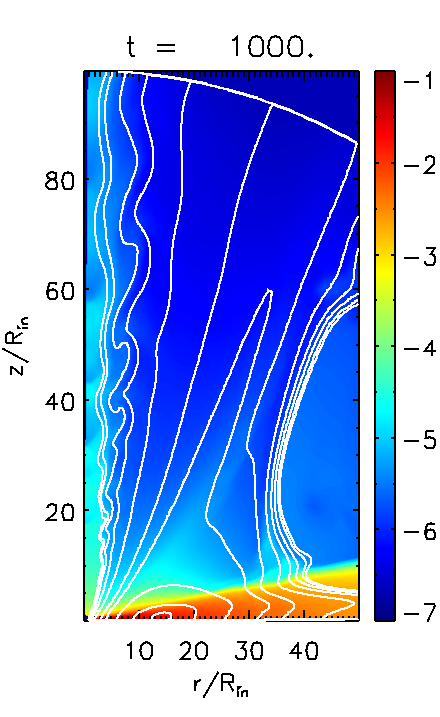}%
\includegraphics[width=0.24\textwidth]{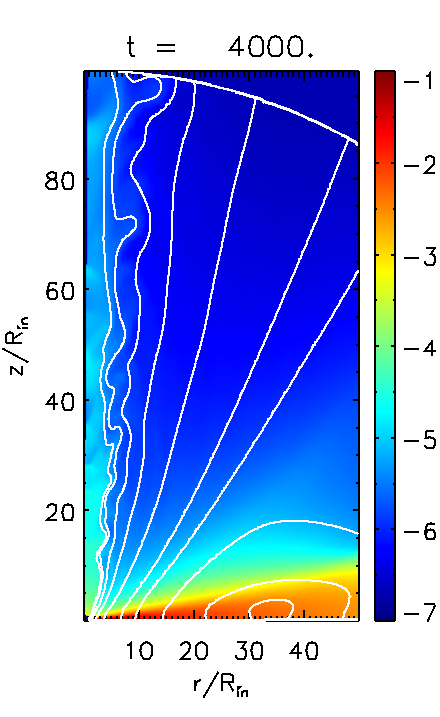}%
\includegraphics[width=0.24\textwidth]{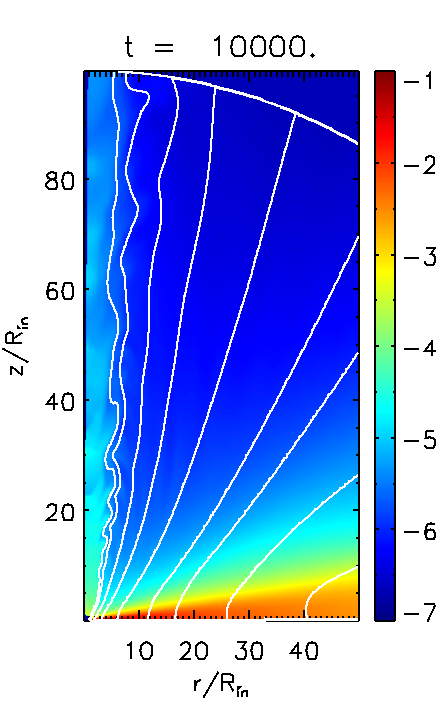}
\caption{Time evolution of the disk-jet structure of the reference dynamo simulation with $\Omega^*=10$.
Shown are simulation steps at $t=[0,1000,4000,10000]$ on a subset of the full numerical grid ($r<50, z<100$).
We display the mass density (colors, in log scale), superimposed by contours of the vector potential,
respectively magnetic flux surfaces.}
\label{fig::ref_evolution}
\end{figure*}

In Figure \ref{fig::ref_evolution} we show the temporal evolution of the reference simulation.
Again the initial setup consists in a weak radial magnetic field confined within the accretion disk.
While the poloidal magnetic field is (if absent, i.e. $B_\theta$) generated and amplified only through a dynamo
effect, the 
toroidal magnetic field is generated by the differential disk rotation and then amplified through the mean-field dynamo.
As discussed in Paper I, the dynamo component $\alpha_\phi$ provides the only mechanism that is
able to amplify the poloidal magnetic field from the toroidal magnetic field.

Essentially, the reference model evolves very similar to the scalar model of Paper I, we hardly 
detect any differences.
The magnetic field is most rapidly amplified in the innermost disk region $t\lesssim500$.
As a consequence, super-Alfv\'enic and super-fast (in the outer domain we reach $v_{\rm jet}\simeq1.5v_{\rm A}$, 
where $v_{\rm A} $)is the Alfv\'en speed) outflows emerge from this part of the accretion disk,
very similar to our toy model and to the literature \citep{2014ApJ...796...29S,2018ApJ...855..130F}, 
while in the outer regions the magnetic field is amplified on a longer timestcale ($t\lesssim5000$).

Also the inclination of the dynamo-generated magnetic field is favorable for the Blandford-Payne magneto-centrifugal 
acceleration mechanism \citep{1982MNRAS.199..883B,1992ApJ...394..117P}, just as in the scalar dynamo simulations.
The jet is ejected from the inner radii of the accretion disk, $R\lesssim10$.  
Its opening angle decreases as it moves away from the disk - thus, the jet becomes collimated.
Because the disk is magnetically diffusive, the magnetic field structure is able to re-arrange, 
leading to a loop structure in the disk without dynamo-inefficient zones (see also Paper I).
This loop structure is swept outward during the long term temporal evolution 
for $t\gtrsim5000$.

\begin{figure}
\centering
\includegraphics[width=0.42\textwidth]{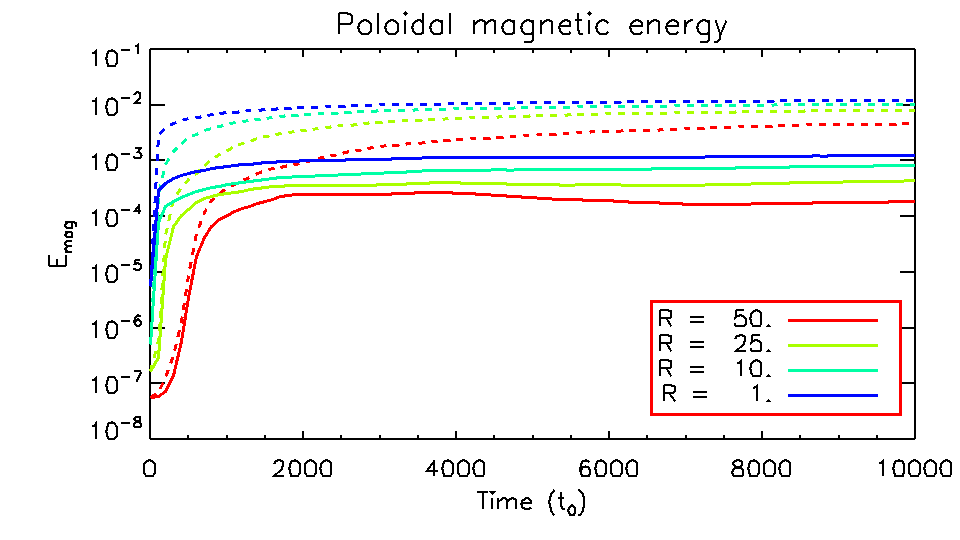}
\caption{Time evolution of the disk magnetic energy for different integration domains for the reference simulation. 
Solid lines show the poloidal magnetic energy, while dashed lines show the total magnetic energy (poloidal + toroidal).
The radii that are labeled denote the lower integration boundary, while the upper integration boundary is at the end of 
the domain, $R = 100$.}
\label{fig::ref_polmag}
\end{figure}

In Fig.~\ref{fig::ref_polmag} we again display the evolution of the disk poloidal and toroidal magnetic energy
as a main signature of the mean-field dynamo, however here derived from a physical model of the dynamo tensor.
The field amplification works on a very short timescales - naturally for a dynamo effect,
with the dynamo working much faster in the inner part of the disk.

After a rapid amplification, the magnetic energy slightly decreases over time.
This is caused by the new model for the dynamo tensor, which now depends on the mid-plane adiabatic sound speed, and therefore is not constant in time.
Although the sound speed shows no significant change through the temporal evolution, it decreases with time due to the mass loss from the disk by accretion and ejection.
We find this behaviour in both scalar and vector dynamo simulations as a consequence of 
the decrease in the dynamo efficiency (sound speed) together with the high diffusivity (diffusive quenching).

\begin{figure}
\centering
\includegraphics[width=0.217\textwidth]{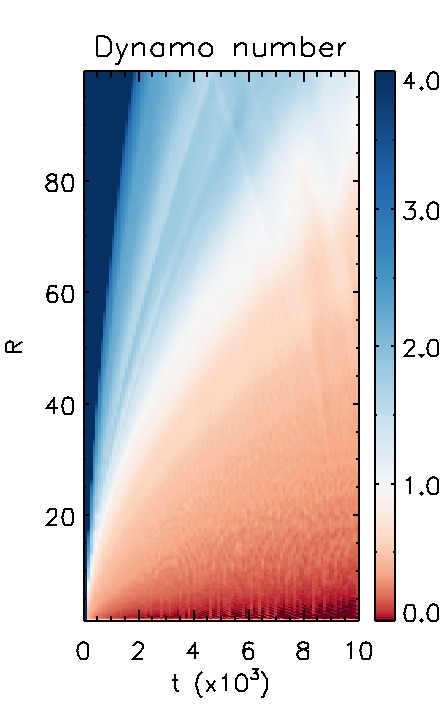}%
\hspace{0.3cm}
\includegraphics[width=0.21\textwidth]{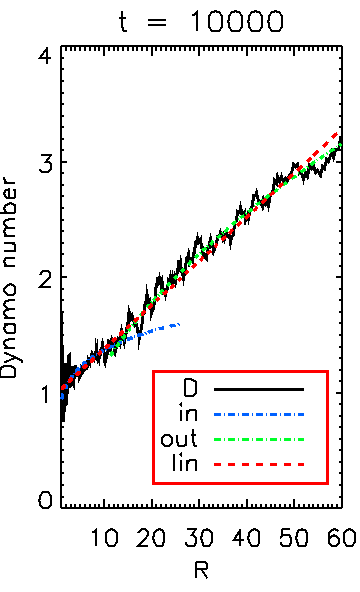}
\caption{Dynamo number $\cal D$ as function of time and radius for the reference simulation. 
The left panel shows the evolution of the dynamo number for all radii. 
The right panel shows the dynamo number at $t = 10000$ within an area of steady state.
The lines denote the dynamo number D (black), and the power law approximations (dashed, see text).}
\label{fig::ref_dynamo}
\end{figure}

As for the toy model, we have considered the dynamo number $\cal{D}$ as a key parameter to determine the 
stability and the evolution of the system (see Fig.~\ref{fig::ref_dynamo}).
In the inner disk region the diffusive quenching acts on a very rapid timescale, saturating the magnetic field and 
decreasing the dynamo number critically below $10$ in the very early stages of the evolution.
As we move further out in radius, the mean-field dynamo leads to a slower and weaker field amplification.
The disk magnetization and, thus, the critical dynamo number, defined as the magnitude of the dynamo number at 
which the disk 
has reached a stable configuration is reached on a longer timescale.
We find that the critical dynamo number is ${\cal D}\simeq10$, which is similar to the magnitude\footnote{The 
critical dynamo number represents the threshold for the onset of non-linear dynamo action. As it depends on the 
physical setup of the problem it is not straight forward to compare these number for different model setups.} 
found in the literature (see e.g. \citealt{2005PhR...417....1B}).

In quasi steady state, the local dynamo number grows with radius (see Fig.~\ref{fig::ref_dynamo}, right panel).
Interestingly, we may fit this dependence with a broken power law.
Thus, after saturation, we may divide the domain of dynamo action into two parts. 
We find an inner part with $R\in[1,20]$ that is best reproduced with a power law exponent $\simeq0.25$, 
while for the outer part for $R > 20$ a square root dependence is the best fit.

As a physical reason for the broken power law we have disentangled the evolution of the 
disk diffusivity, in particular the dependence on the magnetization provided by $\alpha_{ss}$ (see eq. \ref{eq::ssm}).
In the inner region, a power-law approximation of the disk magnetization suggests a power index of $-0.07$ (blue 
dashed dotted line), 
while in the outer region a power index of $-0.17$ is preferred (green dashed dotted line).

Physically, this indicates that the accretion disk is pressure dominated, although very close to a magnetization
constant in radius. 
For this reason, a linear approximation (red dashed line) also provides a reasonable fit good - without the need to 
separate the steady state disk regions into two parts.
Essentially, even if a linear approximation is more simple, the split into two power laws is 
(i) more accurate, and can also be 
(ii) related to the disk physics.

\begin{table}[t]
\caption{Simulations applying the tensor model for the dynamo coefficients.
The sole dynamo parameter is now the Coriolis number $\Omega^{*}$.
The run time of the simulations is $t_{\rm F}$ in units of $1000$.}
\centering
\begin{tabular}{llllll}
  \noalign{\smallskip}
\hline
  \noalign{\smallskip}
 run ID & $\Omega^{*}$ & $ t_{\rm F}$ & Comment\\ 
   \noalign{\smallskip}
 \hline
 \noalign{\smallskip} 
 {\em OM01} &  0.1 & 10 & no jet collimation \\
 {\em OM04} &  0.4 & 10 & dynamo-inefficient zones present \\
 {\em OM1}  &  1.0 & 10 & dynamo-inefficient zones present \\
 {\em OM5}  &  5.0 & 10 & dynamo-inefficient zones absent \\
 {\em OM10} & 10.0 & 10 & reference simulation \\
 \hline
\end{tabular}
\label{tab::tensor}
\end{table}

\subsection{A parameter survey}
\label{sec::param_mag}
In order to understand in more detail how the magnetic field evolution is correlated with a different dynamo 
tensor, we have performed 
simulation runs applying a different Coriolis number $\Omega^{*}$ ranging within $[0,10]$, see Tab.~\ref{tab::tensor}.
We stress again that the Coriolis number compares effects of rotation to those of turbulence, 
with turbulence being responsible 
to amplify a poloidal field while rotation amplifying the toroidal field.

We first have a look at the dynamo coefficients and diffusivity coefficients (see Fig.~\ref{fig::alpha_coeff}).
We see that the $\alpha_z$-component of the dynamo tensor changes sign and is vanishing at $\Omega^* \simeq 1$.
However, this component of the dynamo tensor becomes effectively relevant only for low Coriolis numbers.
This is the limit of low rotation.
In the limit $\Omega^* \to 0$ all the dynamo components tend to vanish, and the magnetic diffusivity becomes isotropic.

\begin{figure}
\centering
\includegraphics[width=0.48\textwidth]{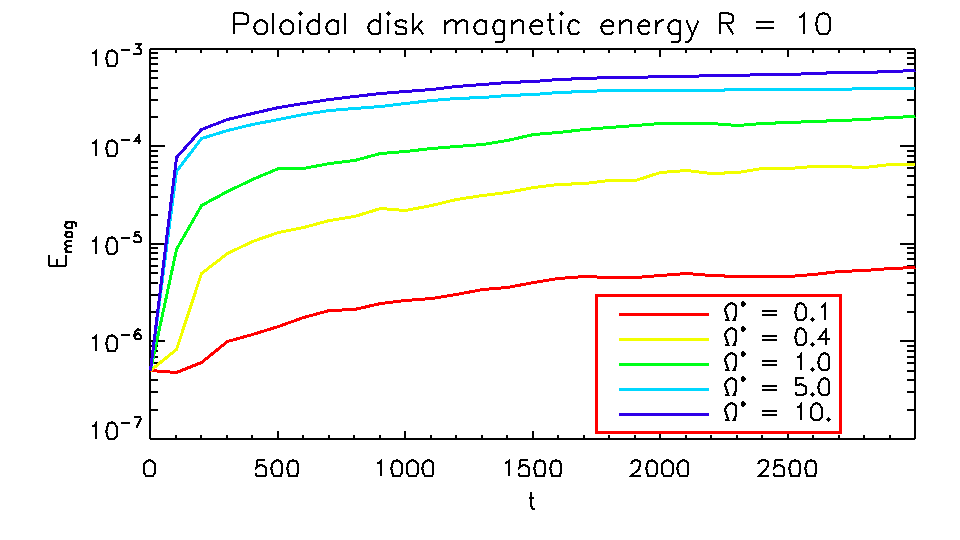}
\includegraphics[width=0.48\textwidth]{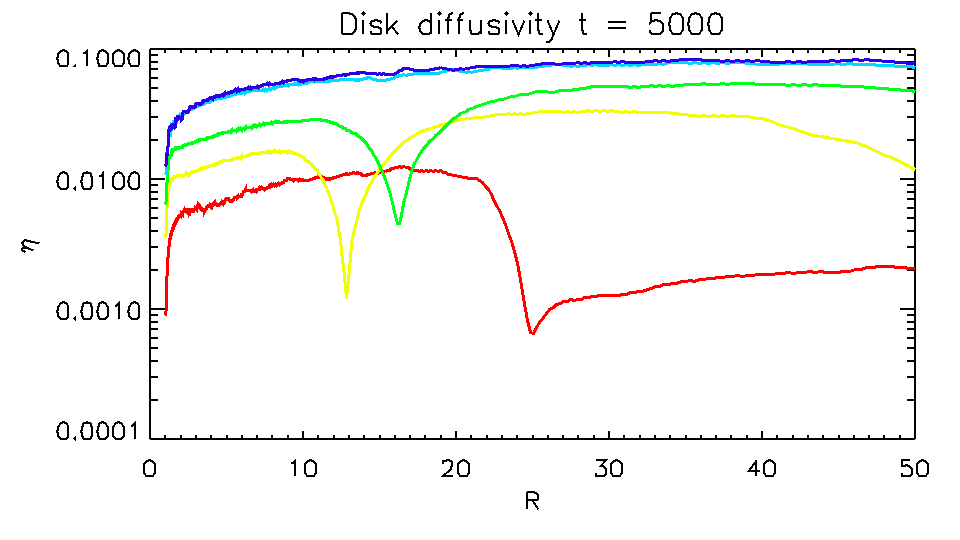}
\includegraphics[width=0.48\textwidth]{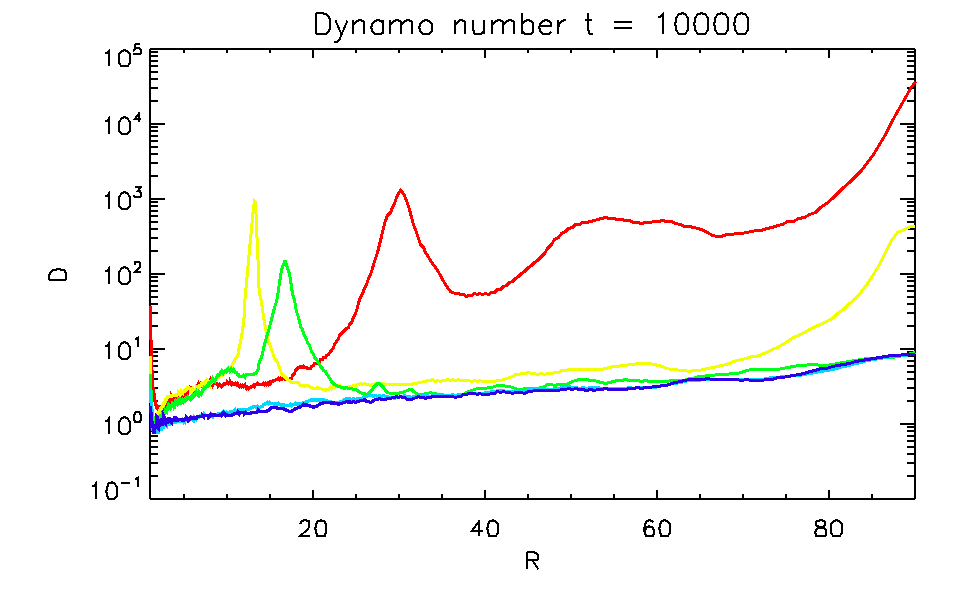}
\caption{Evolution of the magnetic field for different Coriolis numbers $\Omega^*\in[0,10]$.
We show the poloidal magnetic energy (top panel) integrated from $R = 10$ as a function of time, 
the disk diffusivity $\eta$ at $t = 5000$ (middle panel)
and the dynamo number $\cal D$ at $t = 10000$ (bottom panel) as a function of radius along the disk.}
\label{fig::omega_mag}
\end{figure}

\subsubsection{Amplification of the magnetic field}
As for the toy dynamo model, the primary effect of the mean-field dynamo is the amplification of the disk magnetic field.
We first compare the magnetic field amplification for different Coriolis numbers (see Fig.~\ref{fig::omega_mag}).
Since the dynamo component $\alpha_\phi$ depends monotonously on the Coriolis number (see Eq. \ref{eq::alphacoeff}), 
one would expect a higher $\Omega^*$ to result a stronger magnetic field.
However, the critical dynamo number discussed in Paper I is not applicable anymore,
since the Coriolis number has also a strong effect on the disk diffusivity.

What we find is that for $\Omega^* \lesssim 0.15$ the dynamo-amplification of the magnetic is sufficiently efficient 
in order to 
generate a collimated outflow, corresponding to a maximum (absolute) value of $\alpha_{\rm{crit}} \simeq 0.005$.
Note that this value $\simeq10$ times larger than the one recovered by \citet{2018ApJ...855..130F} and almost twice 
as large
as the value that we recovered for our toy model above.

This discrepancy is related to the model for the magnetic diffusivity, which is now self-consistently determined 
by the Coriolis 
numbers, similar to the dynamo-alpha.
In fact, for the critical strength of the dynamo, now also the diffusivity level is higher than in
\citet{2018ApJ...855..130F} and 
and also higher than for the toy model discussed above.
For $\Omega^* \simeq 0.1$, thus slightly below its critical magnitude, the dynamo process is also able to amplify 
the poloidal field, 
however, we do not find collimated outflows from the resulting magnetic field configuration.

We note that a correlation between the profile of disk magnetization and jet collimation has been proposed
already by \citet{2006ApJ...651..272F}, such that a high degree of collimation requires a flat magnetization 
profile, thus a sufficient magnetization also for larger disk radii.
This is what we seem to observe in our dynamo simulations, since the magnetization of case {\em OM01} is lower
for larger radii.

We therefore disentangle the following correlations.
A higher $\Omega^*$ implies a large dynamo efficiency $\alpha_{\phi}$ that leads to a larger disk magnetization 
(stronger field, as the disk gas pressure remains similar), which finally supports jet collimation.
For $\alpha_\phi \gtrsim \alpha_{\rm{crit}} \simeq 0.005$ the poloidal magnetic field is amplified to different 
magnitudes and also on different timescales.
Naturally, a stronger dynamo term, as shown in Paper I, leads to a stronger amplification of the 
poloidal magnetic field on a faster timescale.
In particular we see that the poloidal magnetic energy increases rapidly before $t = 500$, and after a strong
amplification, the saturation state is reached on a later timescale. 

Since a weaker dynamo can amplify the poloidal magnetic field only to lower strength, the poloidal disk magnetic 
energy does not increase immediately in the case of $\Omega^* \simeq 0.1$.
This is simply due to the evolution of the magnetic diffusivity, which follows a faster timescale than the
dynamo-$\alpha_\phi$.
However, since the toroidal field is amplified from the initial field by the $\Omega$-effect, the poloidal field is
eventually amplified as well.

\subsubsection{Magnetic diffusivity and dynamo number}
\label{sec::omega_diffusivity}
We now investigate how the magnetic diffusivity and the dynamo number evolve with respect to our main simulation 
parameter, the Coriolis number.
In Figure~\ref{fig::omega_mag} (middle panel) we show the disk magnetic diffusivity profile for different Coriolis 
numbers at $t= 5000$.
We may identify three different evolutionary characteristics.

For (i) high Coriolis numbers, $\Omega^* \gtrsim3$, the diffusivity profile is very similar to the one for the 
reference simulation
with $\Omega^* = 10$ (blue curve).
The diffusivity profile remains somewhat constant for $10^{-2} < \eta < 10^{-1}$. 
Here, the magnetic field amplification leads to an increase of diffusivity quite rapidly (diffusive quenching)
and a steady state is reached soon at $t\lesssim500$ in the inner disk region.

For (ii) lower Coriolis number dynamo-inefficient zones are formed (one or more) within the accretion disk, due
to the low $\alpha_R$.
These dynamo-inefficient zones are clearly visible in Fig.~\ref{fig::omega_mag} as zones where the magnetic disk
diffusivity sharply decreases.
This behavior can be seen for simulations applying $0.4 < \Omega^* < 1.0$.

For (iii) even lower Coriolis numbers, e.g. for $\Omega^* = 0.1$, the magnetic field amplification remains low.
Therefore, in addition to the emerging magnetic loops, the dynamo in outer regions of the disk is not able to 
amplify the magnetic field.
Again, as discussed above, because of the weak magnetic field, magnetic diffusivity remains low as well.
Still, the inner disk has a substantial magnetic field and also a high diffusivity. 

In order to understand if and where the amplification of the magnetic field is saturating, we have a 
look at the dynamo number at $t = 10000$ (Fig.~\ref{fig::omega_mag}, lower panel).
For larger $\Omega^*$, e.g. $\Omega^*\gtrsim3$, the magnetic field (both poloidal and toroidal) has been amplified 
in all areas of the accretion disk at this time (but not in the dynamo-inefficient zones).
As we know, the actual amplification of the magnetic field plays a key role in the diffusive quenching model 
(see Eq.~\ref{eq::ssm}).
Therefore, for the Coriolis numbers considered, the dynamo number, which directly depends on the magnetic 
diffusivity, falls under a critical magnitude for dynamo action.

This does not apply for the dynamo-inefficient zones.
Although these zones are characterized by a large dynamo number, they are not correlated with the amplification of the magnetic field.
With a lower Coriolis number, the magnetic field amplification occurs on longer timescales, especially for the
outer disk.
For this reason, besides the dynamo-inefficient zones, the dynamo number remains over its critical magnitude 
also in the outer disk regions,
for which just more time would be required in order to reach a magnetic field saturation.
Moreover, for $\Omega^*\lesssim0.1$, the dynamo number is 
not a good measure for the mean-field dynamo, since it is not connected anymore to the process of field amplification.


\subsection{Dependence on the initial seed field}
Mean-field dynamo action is expected to be independent on the initial seed field, due to the exponential growth 
by the dynamo amplification.
However, we discovered that second-order effect of the initial evolution may affect also the long term evolution 
of the system.

In Paper I we have discussed the impact of the dynamo component $\alpha_\theta$ in the toy model.
We had found that when applying a vertical {\em initial} magnetic field, the scalar dynamo model may lead to a 
non-physical hydrodynamical evolution, mainly caused by low density zones forming in the proximity of the inner radial boundary.
The origin of these numerical issues seems to be due to the formation of dynamo-inefficient zones in the very inner part
of the accretion disk.
Since for the toy model there are no {\it a priori} constraints on the dynamo tensor components, we also have 
tested the effects of an initial vertical seed field with a reduced strength of $\alpha_\theta$ ($\psi = 0.1$), 
just in order to avoid the formation of the dynamo-inefficient zones in the inner disk.

\begin{figure}
\centering
\includegraphics[width=0.24\textwidth]{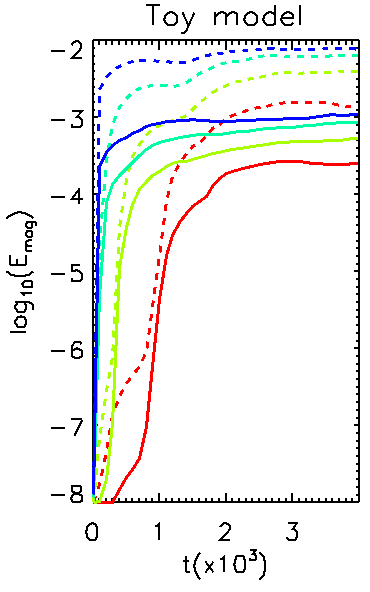}%
\includegraphics[width=0.24\textwidth]{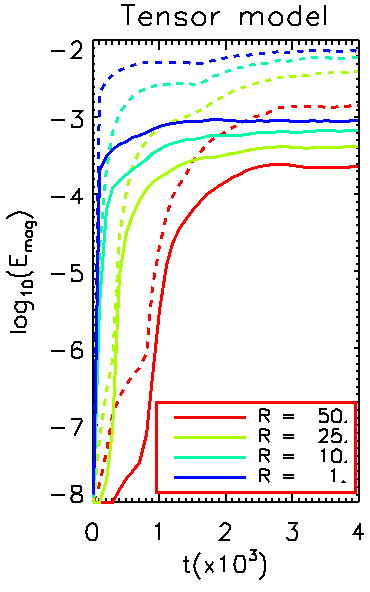}
\caption{Time evolution of the disk magnetic energy for simulations applying a vertical seed field.
The  radii  that  are labeled  denote  the  lower  integration  boundary,  while  the  upper integration boundary 
is at the end of the domain, $R= 100$.
Solid lines denote the poloidal magnetic energy, while dashed lines show the total magnetic energy
(poloidal + toroidal). }
\label{fig::ref_polvert}
\end{figure}

In the analytical model of \citet{1995A&A...298..934R} the anisotropy of the tensor component $\alpha_\theta$ is 
introduced naturally on physical grounds and it does not require any additional constraint.
We have performed a simulation with $\Omega^* = 10$ and a vertical initial magnetic field (applying a vector
potential $A_\phi = 10^{-5}$.
Indeed, the results are comparable with the simulations run th\_B of Paper I (see Fig.~\ref{fig::ref_polvert}).

Here the component $\alpha_\theta$ is suppressed, as directly inferred from analytical dynamo theory, and no 
ad-hoc assumption of anisotropy is required.
Therefore, the effects of shear between the rotating disk and the steady-state corona are not amplified by the 
dynamo as they were in the scalar dynamo model.

As demonstrated in Paper 1, the amplification of the poloidal disk magnetic field occurs on different 
time scales depending on the distance from the central object.
Although during early stages the field amplification looks to the case of an initially radial initial field 
(see e.g. Fig.~\ref{fig::ref_polmag} for a comparison), at $t = 4000$ the poloidal magnetic energy that is 
dynamo-amplified is comparable.

The saturation of the magnetic field amplification towards the same magnitude is evidence for the ongoing action 
of the mean-field dynamo, 
which is able to generate a magnetic field regardless of the initial magnetic field configuration.
The fact that the two panels of Fig.~\ref{fig::ref_polvert} are basically indistinguishable from 
Fig.~\ref{fig::ref_polmag} indicates how much the component $\alpha_\theta$ is overestimated in the scalar 
dynamo model when non-radial initial magnetic field is present. 
This is a clear advantage of the tensor model, since it allows to suppress the different dynamo components without 
adding additional constraints.

A substantial difference between simulations applying an initially radial or vertical initial field, respectively,
is the formation of dynamo-inefficient zones even for $\Omega^*=10$.
This implies that anti-aligned magnetic loops can form also 
in case of a high Coriolis number.

Overall, the evolution of dynamo-inefficient zones may also depend on the quenching model and the diffusivity model.

\subsection{Accretion and ejection}
A difference in the magnetic field structure plays a key role in the dynamics of the accretion disk and the outflow.
This holds for the toy model for the dynamo tensor as well as for the physical model for the tensor components.
In this section we want to discuss the dynamical evolution of the accretion-ejection structure for the model of
\citet{1995A&A...298..934R} and compare the results for different Coriolis numbers $\Omega^*$.

In fact, as a first general result we do not significant differences between the scalar toy model and the reference 
simulation {\em OM10}.
This nice agreement validates the model approach described in Paper I in the context of jet 
launching large scale simulations.

\begin{figure}
\centering
\includegraphics[width=0.48\textwidth]{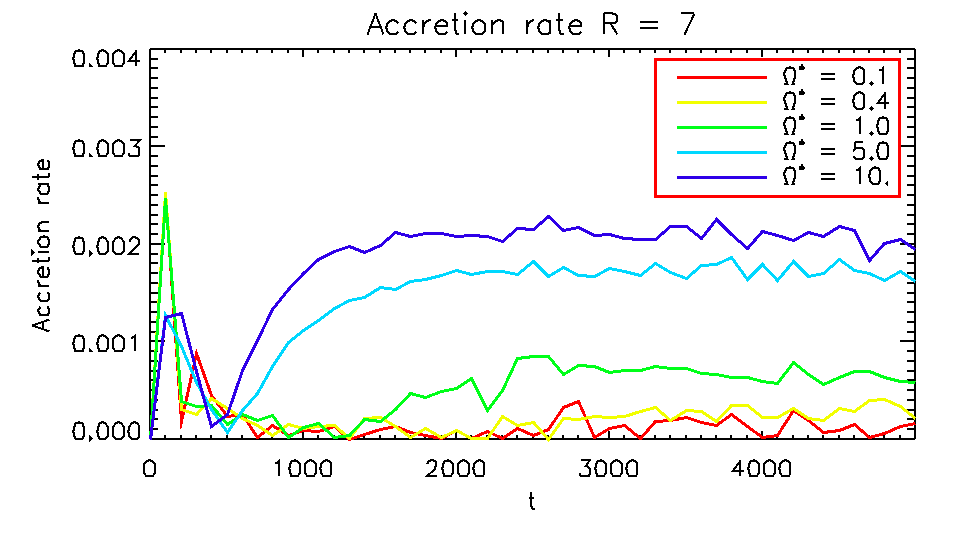}
\includegraphics[width=0.48\textwidth]{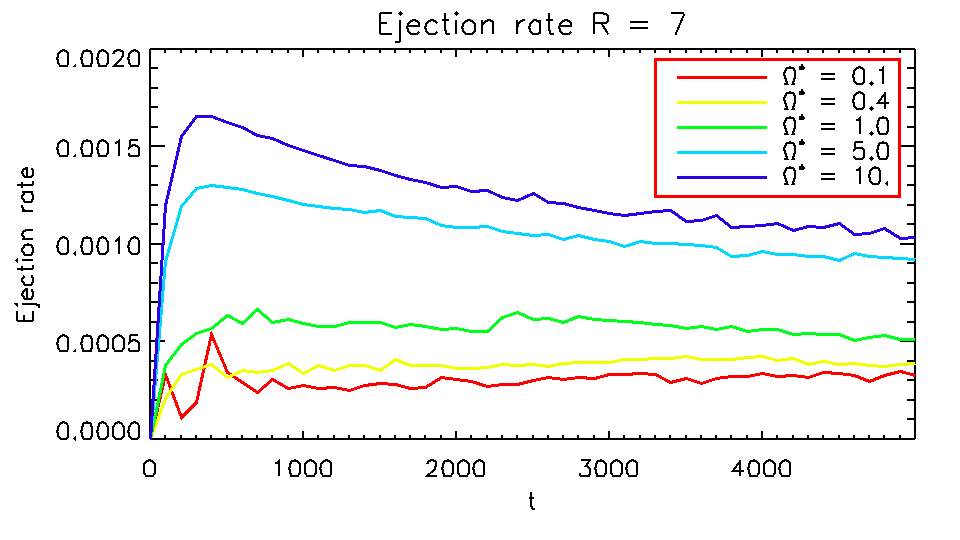}
\caption{Evolution of the accretion (top panel) and ejection (bottom panel) rates for different 
Coriolis numbers $\Omega^*\in[0,10]$. 
The accretion rate is computed at $R = 7$, while the ejection rate is computed along the disk surface between 
$R = 1$ and $R = 7$
(see Paper I, appendix for a definition of the control volume).}
\label{fig::omega_accr_ej}
\end{figure}

We now compare further simulation runs.
We first consider the accretion and ejection rates in Fig.~\ref{fig::omega_accr_ej}.
The accretion rate increases with the Coriolis number, meaning it increases as well with the strength of the 
mean-field dynamo.
This is because a stronger field amplification, implying a higher disk magnetization, leads to a higher diffusivity
and therefore facilitates accretion.
In addition, a stronger magnetic field is also more efficient in angular momentum removal.
When dynamo-inefficient zones are present (see Fig.~\ref{fig::omega_mag}), they effectively enhance the difference
between accretion and ejection rates as we have discussed already in Paper I.

The ejection rate, increases with the Coriolis number, similar to the accretion rate.
In general, the ejection-accretion ratio is higher for a lower dynamo efficiency, in agreement with previous
simulations \citep{2014ApJ...796...29S} and with 
the toy dynamo model, as it depends on the dynamo components $\alpha_\phi$ and $\alpha_R$.

We also notice a slow decrease over time in the ejection rates, which we understand are due to subtle changes in the disk dynamics.
Such variations could be triggered by the disk mass loss, which in turn effects the dynamo tensor components, as
they are parameterized by the sound speed at the disk mid-plane.

Before reaching the quasi-steady state, the accretion-ejection rate, defined as $\dot{M}_{\rm{eje}}/\dot{M}_{\rm{acc}}$ 
(see appendix Paper I), may exceed unity\footnote{This is
impossible in steady-state, as the disk mass will be dispersed rapidly}.
The reason of such a high ejection efficiency in early evolutionary stages is due to the time scales of the 
processes involved.
In fact, accretion requires more time to establish and to saturate, while ejection operates on a faster timescale.

A reason why there evolves a more turbulent state of the accretion disk,
is the magnitude of $\alpha_R$, which changes as well with the Coriolis number.
As shown before, for a lower strength of $\Omega^*$ magnetic loops are formed in the disk, implying a more turbulent evolution.
A peculiar case is when $\alpha_\phi < \alpha_{\rm crit}$ (e.g. for {\em OM01}).
Here, the magnetic field is amplified, but not to a sufficient strength in order to collimate the jet.
In this case the accretion rate -- correlated to the magnetic diffusivity -- is almost negligible,
however, we still find some slight ejection in the form of un-collimated disk winds. 

\begin{figure}
\centering
\includegraphics[width=0.48\textwidth]{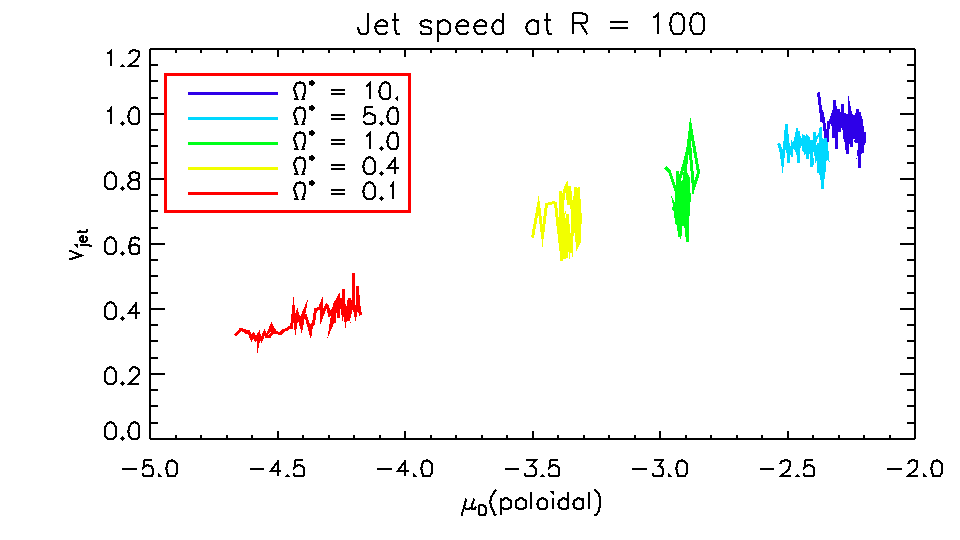}
\caption{Jet speed vs disk magnetization. 
Shown is the maximum jet velocity versus the disk magnetization calculated from the poloidal magnetic field for different 
Coriolis numbers $\Omega^*\in[0,10]$.}
\label{fig::omega_velmag}
\end{figure}

The differences in the mass loading and in the magnetic field reflect on the jet speed and kinematics.
As for the toy model (see Paper I) we expect the jet speed increase with the magnetization, which is strictly correlated
with the Coriolis number $\Omega^*$.

The correlation between poloidal disk magnetization and jet speed is shown in Fig. 
\ref{fig::omega_velmag}.
The increasing in the jet speed as a function of the disk magnetization shows a nice agreement with \citet{2016ApJ...825...14S} and with the toy model.
We find that for $\Omega^*\gtrsim1$ the jet speed reaches the Keplerian velocity at the inner disk radius, which is a well-know result
for jet formation simulations (see e.g. \citealt{1997ApJ...482..712O, 1999ApJ...526..631K}), and decreases for lower values of the Coriolis number.

\begin{figure*}
\centering
\includegraphics[width=0.19\textwidth]{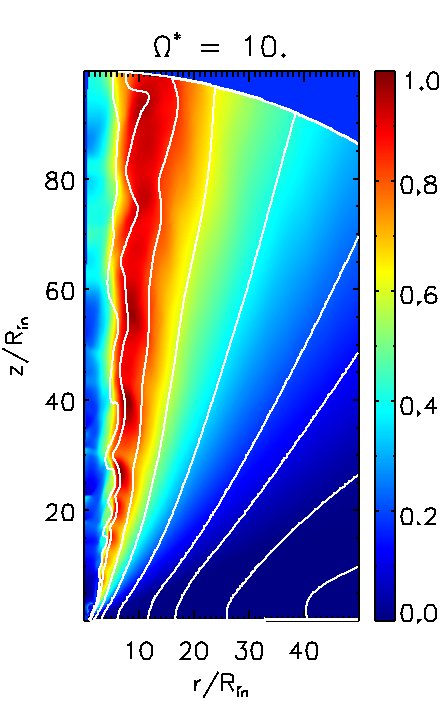}%
\includegraphics[width=0.19\textwidth]{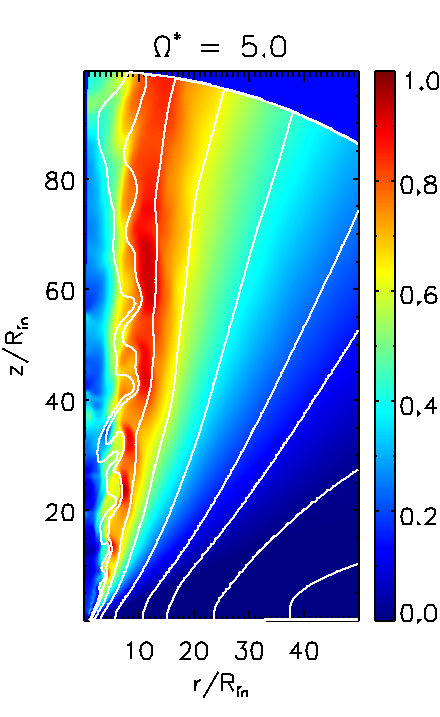}%
\includegraphics[width=0.19\textwidth]{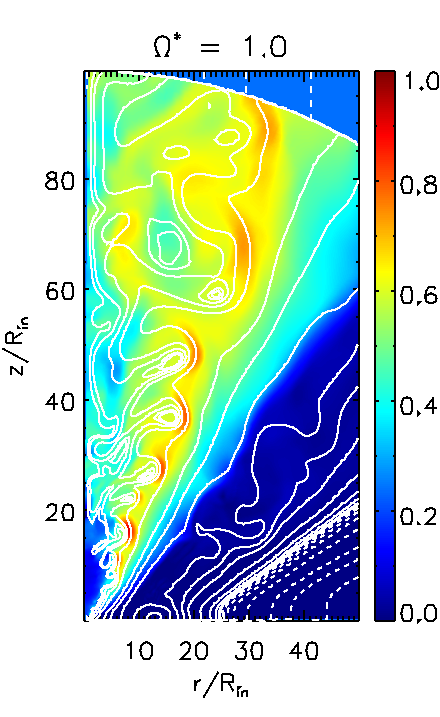}%
\includegraphics[width=0.19\textwidth]{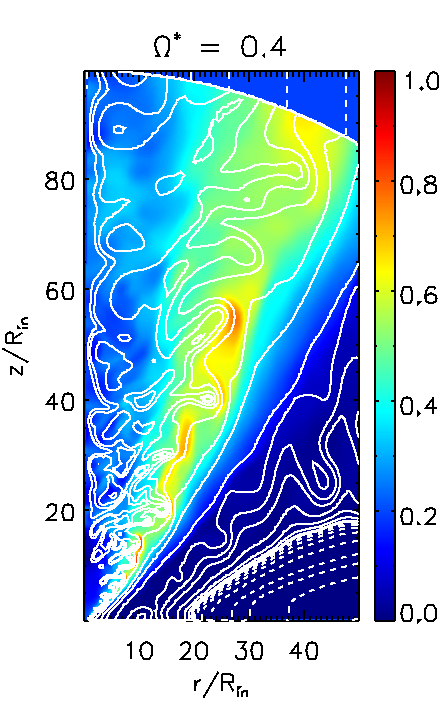}%
\includegraphics[width=0.19\textwidth]{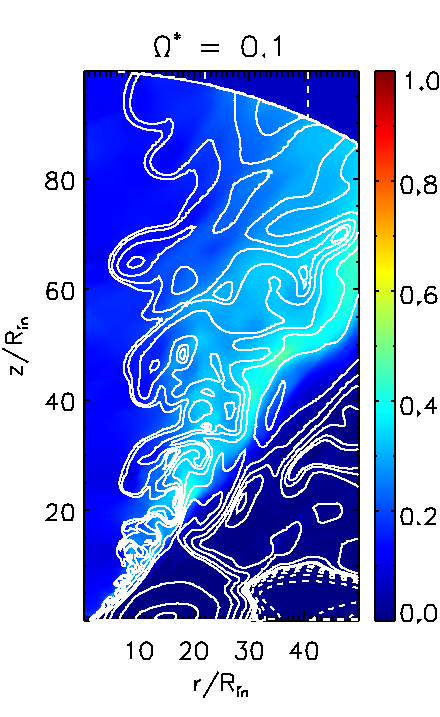}
\includegraphics[width=0.19\textwidth]{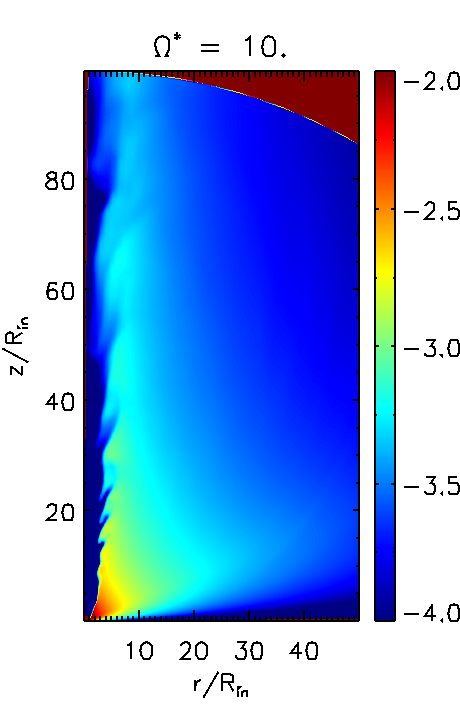}%
\includegraphics[width=0.19\textwidth]{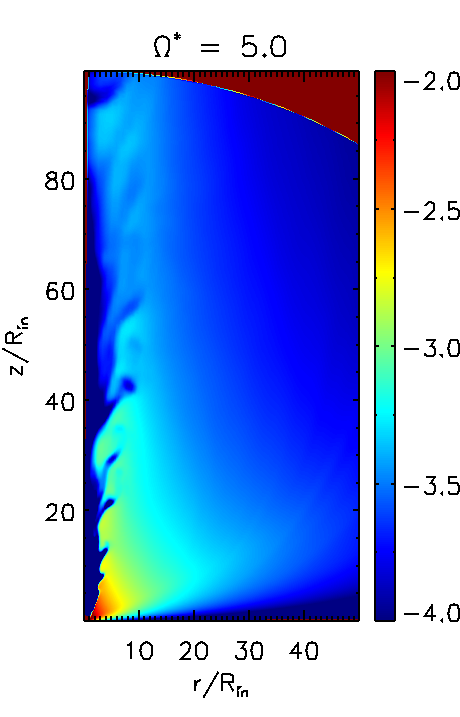}%
\includegraphics[width=0.19\textwidth]{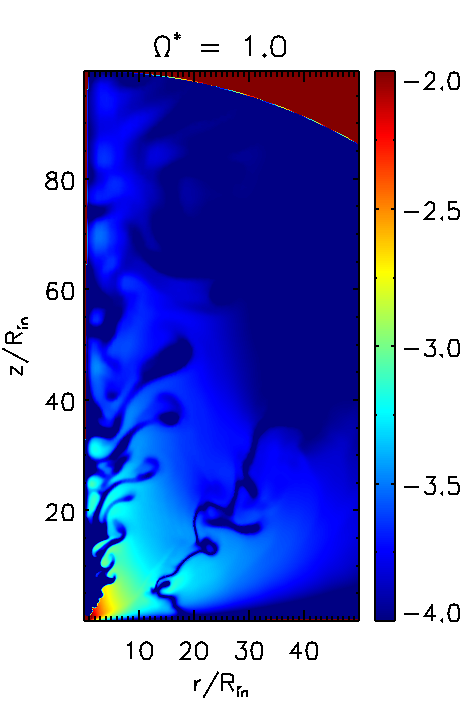}%
\includegraphics[width=0.19\textwidth]{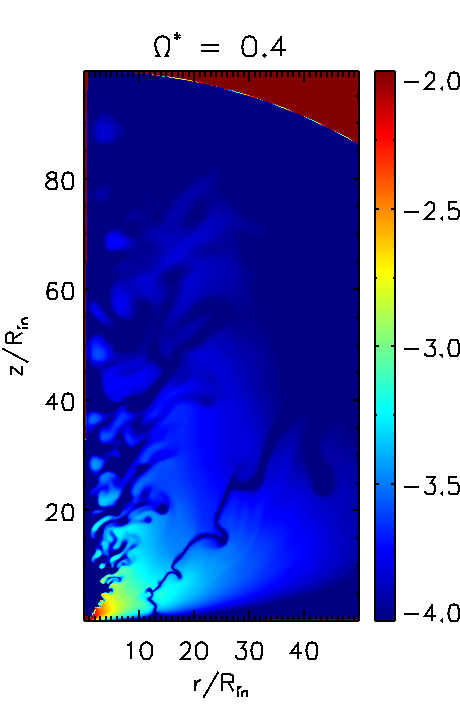}%
\includegraphics[width=0.19\textwidth]{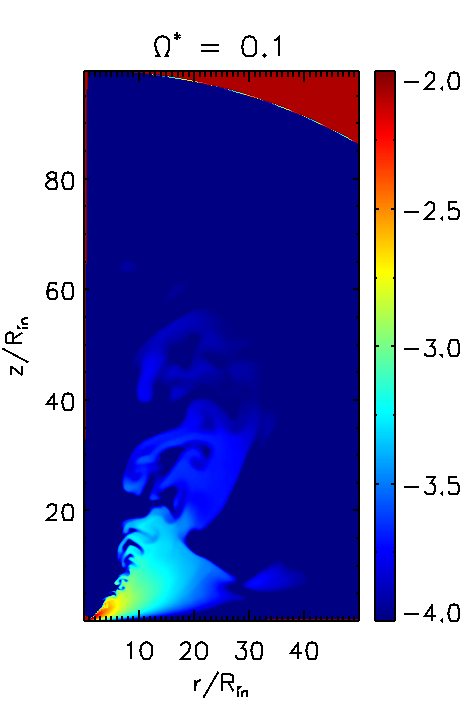}
\caption{Comparison of parameter runs at $t = 10000$.
 Shown is the distributions of the poloidal velocity (top), overlaid with contour lines of the vector 
 potential (following poloidal field lines)(top), and toroidal magnetic field strength (bottom) for different values of the Coriolis number $\Omega^*$.}
\label{fig::omega_jet}
\end{figure*}

Another observable is the jet collimation, which shows the impact of the disk dynamo on the jet dynamics.
Using the same definition of collimation used in Paper I, we see from Fig.~\ref{fig::omega_jet}
how the Coriolis number (and therefore the dynamo tensor) affects the jet collimation.
As shown in Section \ref{sec::omega_diffusivity}, it is possible to find three different outcomes.
For high Coriolis number ($\Omega^*\gtrsim3$), we find a highly collimated jet.
For $\Omega^*\lesssim3$ the evolution is characterized by the formation of dynamo inefficient zones,
which play a key role in the jet speed and collimation.
The structure of the poloidal magnetic field is more turbulent, which implies a less collimated jet.
In addition, a lower value of the Coriolis number means also a weaker $\alpha_\phi$ component, which 
leads to a weaker disk magnetization (see Fig.~\ref{fig::omega_velmag}) and therefore, 
in agreement with \citet{2006ApJ...651..272F}, a less collimated jet.
Below the critical Coriolis number ($\Omega^*<0.15$) the amplification of the poloidal field does not occur,
and therefore the outflow is not collimated.
We also see that the toroidal field is not able to expand through the domain,
and it remains confined in the inner regions of our domain.
This results are a combination of the two main results found in Paper I, i.e. the strength of the component
$\alpha_\phi$ and the formation of the dynamo inefficient zones.

Here we may close the loop to the observed jet quantities.
Overall we find that magnetic fields generated by a disk dynamo can well launch outflows and accelerate and collimate 
them into jets. 
In particular this holds for a anisotropic dynamo of a thin disk, which can produce a disk magnetization that is able
to eject strong jets. 

However, we  also find that in other that than {\em thin} accretion disks the dynamo is influenced also by other tensor
components.
Those lead to more unstable, more structured, but slower outflows, which may potentially not survive on the 
observed spatial scales. 
We find a variation in the jet speed between 0.3 and 1.1 the Keplerian speed at the inner disk orbit.

We propose that the variety of observed jet structures thus may reflect the underlying variation of accretion disks, 
both coupled by the disk-dynamo generated magnetic field.


\section{Conclusions}
\label{Sec:conclusions}
We have presented MHD dynamo simulations in the context of large-scale jet launching.
Essentially, a magnetic field that is amplified by a mean-field disk dynamo, is able to drive a high speed jet.
All simulations have been performed in axisymmetry, treating all three vector components for the magnetic field 
and velocity.
We have applied the resistive code PLUTO 4.3 \citep{2007ApJS..170..228M}, 
however extended by implementing an additional term in the induction equation
that considers the mean-field dynamo action.

Extending our approach from Paper I where we applied (ad-hoc) choices for the dynamo tensor components,
here we consider an analytical model of turbulent dynamo theory \citep{1995A&A...298..934R} that incorporates
both the magnetic diffusivity and the turbulent dynamo term, connecting their module and anisotropy by only one 
parameter, the Coriolis number $\Omega^*$.

In particular we have obtained the following results:

1) 
The prime advantage of the tensor dynamo model is the reduced number of the 
parameter space, in combination with the physically more consistent approach for the dynamo.
Both the dynamo and the diffusivity tensor can be fully recovered from one single parameter -- the Coriolis number $\Omega^*$.
Another significant advantage of the tensor model is the physical constraint for the different dynamo components.
Applying a non-radial seed magnetic field, the tensor model naturally suppresses the dynamo action by 
the component $\alpha_\theta$, which plays a key role in presence of a non-radial initial magnetic field. 

2)
Our new approach confirms the previous results of dynamo simulations, as they are included in the new modeling 
as a limiting case (e.g. \citealt{2014ApJ...796...29S,2018ApJ...855..130F}).
Essentially, the tensor dynamo model shows very good agreement with previous studies and the toy model described 
in Paper I, recovering very similar results, thereby approving the approach of the toy model. 
Looking at different Coriolis numbers, we can distinguish between high values ($\Omega^*\gtrsim3$),
where the disk shows no dynamo-inefficient zones, a low $\Omega^*\lesssim3$, where the evolution of the 
disk is affected by the formation of one or more dynamo-inefficient zones.
For even lower $\Omega^*\lesssim0.15$ dynamo-inefficient zones form and the disk magnetization does not saturate 
at large radii -- both effects affect the jet collimation on the simulation time scales considered.

3) We have studied the evolution of the launching process and and also the properties of the ejected jet flow 
   the for different Coriolis numbers $\Omega^*$ that affect the dynamo process.
  We find that a higher $\Omega^*$ leads to a stronger amplification of the magnetic field. 
  This results is in agreement with previous (scalar) mean-field dynamo simulations, but is now put on a more 
  physical ground as it is connected to a more physical disk dynamo model.

4) We have further extended the correlation found by \citet{2016ApJ...825...14S} and in Paper I between the accretion disk 
   magnetization and the jet speed, linking the former quantity to the mean-field dynamo. 
   In particular we have found that higher values of the Coriolis number $\Omega^*$ lead to a stronger magnetization within the 
   accretion disk and therefore to a faster jet.
   If the Coriolis number (and therefore the dynamo) is not strong enough to amplify the poloidal magnetic field, we 
   find an uncollimated outflow in form of slow disk wind.

5) We have investigated the formation of the so-called {\em dynamo-inefficient zones}
   for different values of the Coriolis number and their effect on the disk-jet connection. 
   We find that for small Coriolis numbers $\Omega^*\lesssim3$, dynamo-inefficient zones are formed in the accretion disk.

6) We have investigated the detailed physical interaction of the dynamo with the field structure by applying a
   vertical seed magnetic field following 
   the initial evolution  of the field amplification by the dynamo tensor component $\alpha_\theta$, which is
   naturally overestimated in the scalar dynamo 
   model (for disk dynamos).
   Essentially, we find that a non-isotropic dynamo leads to more stable evolution of the disk-jet system, 
   since the component $\alpha_\theta$ (leading to 
   a magnetic field sub-structure)  is naturally suppressed without any additional constraints.
   
7) We finally emphasize the astrophysical relevance of our findings. 
   Firstly, dynamo generated magnetic fields can well launch outflows and accelerate and collimate them into jets. 
   This holds in particular for a turbulent, anisotropic disk dynamo, which can produce strong jets. 
    Secondly, other than thin accretion disks are influenced also by other dynamo tensor components that lead to 
    more unstable, more structured, but slower outflows, which may potentially not survive on the observed spatial scales. 
    We find a variation in the jet speed between 0.3 and 1.1 the Keplerian speed at the inner disk orbit.
    Thirdly, the observed variety of jet structures thus may reflect the underlying variety of accretion disks, that is coupled 
    to the outflows via the disk-dynamo generated magnetic field.
   
So far we have not looked for unsteady jet launching process, which can lead to the pulsed ejection that is 
observed in most jet sources.
The model investigated does not have any direct feedback of the magnetic field on the dynamo term. 
Future simulations should include more physical feedback (e.g. quenching) models and will presented in a 
forthcoming paper.


\acknowledgements
We thank Andrea Mignone and the PLUTO team for the possibility to use their code.
All the simulations were performed on the ISAAC cluster of the Max Planck Institute for Astronomy.
We acknowledge many helpful comments by an unknown referee that lead to a clearer structure 
of our paper.

\appendix

\section{Resolution study}
\label{sec::resolution}
\begin{figure*}
\centering
\includegraphics[width=0.19\textwidth]{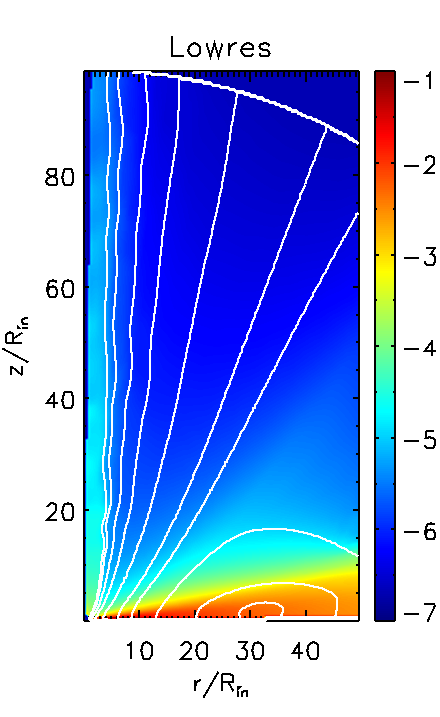}%
\includegraphics[width=0.19\textwidth]{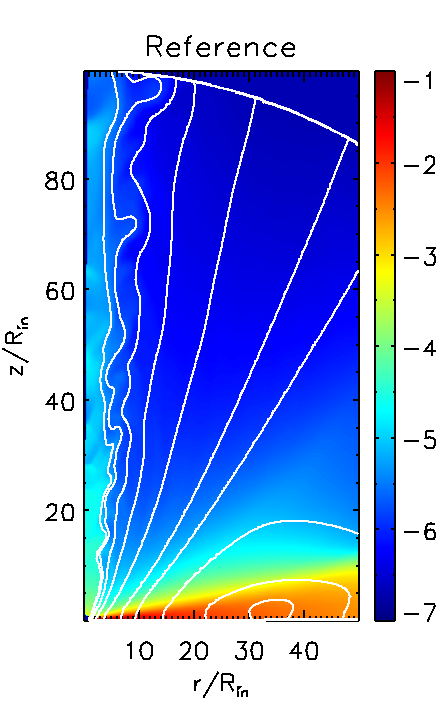}%
\includegraphics[width=0.19\textwidth]{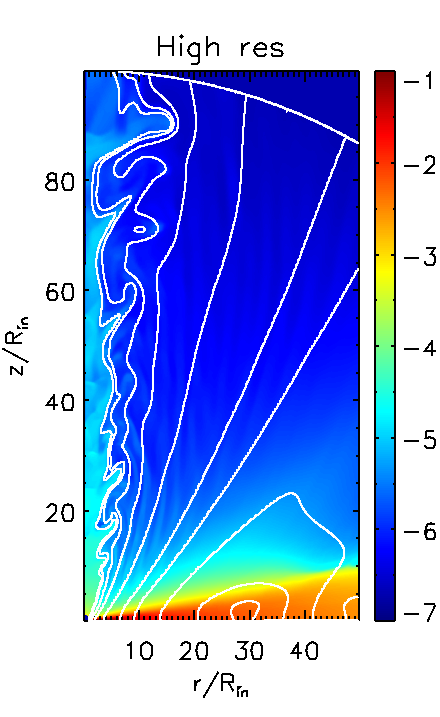}
\includegraphics[width=0.19\textwidth]{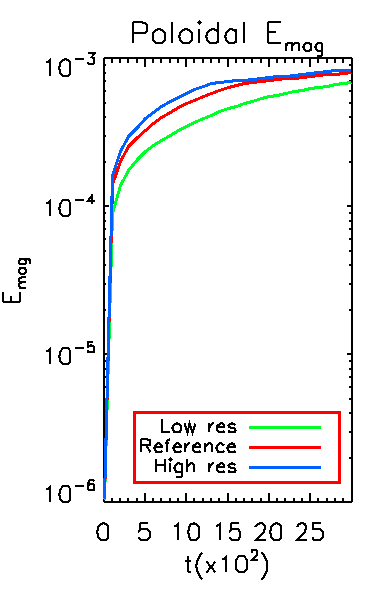}%
\includegraphics[width=0.19\textwidth]{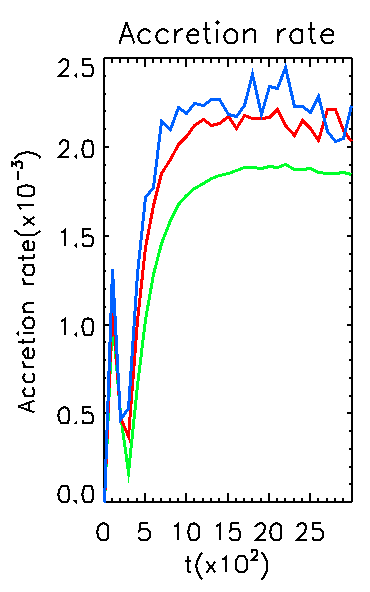}
\caption{Resolution study. 
Density distribution (color) and poloidal magnetic field (white lines) at $t = 4000$ for simulations applying the reference
parameters, but for different resolution (three left panels).
temporal evolution of the poloidal disk magnetic energy ($4^{\rm{th}}$ panel) integrated from $R = 5$ to the outer 
boundary, $R = 100$,and the accretion rate (right panel) computed at $R = 5$.}
\label{fig:res_density_polmag}
\end{figure*}

A numerical study is incomplete without presenting a resolution study.
This is done the in following where we discuss how our physical results depend on the numerical resolution applied.
We compare our reference simulation (resolution $[512\times128]$) of the tensor model (Section \ref{sec::ref})
with two simulation runs applying exactly the same physical parameters, but different resolution.
We choose $[1024\times512]$ for a higher resolution run and $[256\times64]$ for a lower resolution run.
The results are displayed in Fig.~\ref{fig:res_density_polmag} where we show the density and poloidal magnetic
field distribution and the evolution of the dynamo-generated poloidal magnetic energy.

First of all we notice that the reference resolution shows very small differences with the high resolution case, and this 
mostly in the initial evolutionary stages.
The open field lines, favorable for the launching, in the inner disk region and the magnetic loops in the outer 
disk are present in all simulations, 
with almost no difference (see Fig.~\ref{fig:res_density_polmag}).
This holds in particular for the evolution of the disk poloidal magnetic energy.
On the other hand, for the low resolution run the differences persist also on the later stages, although the 
qualitative temporal evolution is the same of the reference case (see Fig.~\ref{fig:res_density_polmag}).

The differences in the evolution of the magnetic field are mostly related to the different numerical diffusivity, 
which is higher for lower resolution.
Before the dynamo quenching by diffusivity has taken place, we believe that the numerical diffusivity quite 
contributes in the low resolution 
case, leading to a damping of the magnetic field amplification (a higher diffusivity lowers the dynamo number).
However, at later times the {\em physical} magnetic diffusivity (which is triggered by the disk magnetization)
becomes dominant and 
therefore the poloidal magnetic energy saturates around the same level (see Fig.~\ref{fig:res_density_polmag}).

Numerical diffusivity plays a key role in the dynamics of the disk-jet connection, e.g. in the efficiency of the 
accretion process, and also for the
mass loading of the disk wind.
Since in the low resolution case the field amplification slower, the saturation of the diffusivity level that 
allows to replenish (by accretion) the
disk matter from the outer disk, happens on a longer timescale as well.
Therefore, the disk accretion rate decreases for the lower resolution setup.

In summary, our simulation results are not completely resolution independent.
However, the results of our reference simulation are very close to a higher resolution study, so a higher resolution 
would not lead to any improvement.
In contrary, a lower resolution would affect the hydrodynamics of the system as well as the evolution of the 
magnetic field.
Thus, we conclude that the resolution we chosen is in fact appropriate in order to capture the essential physics 
while keeping the computational low.

%
%
%
\vspace{2mm}
\bibliographystyle{apj}

\end{document}